\documentclass[twocolumn]{aastex62}
\usepackage{mathtools}
\usepackage[]{algorithm}
\usepackage{algpseudocode}
\usepackage{fontawesome}
\usepackage{soul}

\newcommand\given[2]{\left(#1\;\middle|\;#2\right)}

\newcommand{\boldSigma}{\boldsymbol{\Sigma}}
\newcommand{\boldSigmaErr}{\boldsymbol{\Sigma}_{\rm err}}

\newcommand{\boldX}{\boldsymbol{x}}

\newcommand\hatbolds{\widehat{\boldsymbol{s}}}
\newcommand\bolds{\boldsymbol{s}}
\newcommand\boldexps{\langle\boldsymbol{s}\rangle}

\newcommand{\boldcorr}{\boldsymbol{R}}
\newcommand{\boldvar}{{\rm diag}(\boldsymbol{\tau})}

\newcommand{\sigmagp}{\sigma_{gp}}

\shorttitle{Population Profile Estimator (PoPE)}
\shortauthors{Farahi, Nagai \& Chen}

\begin{document}
\title{\textsc{PoPE}: A population-based approach to model spatial structure of astronomical systems}
\correspondingauthor{Arya Farahi}\email{aryaf@umich.edu}

\author[0000-0003-0777-4618]{Arya~Farahi}
\affil{The Michigan Institute for Data Science, University of Michigan, Ann Arbor, MI 48109, US}
\affil{Department of Physics and Leinweber Center for Theoretical Physics, University of Michigan, Ann Arbor, MI 48109, US}
\author{Daisuke~Nagai}
\affil{Department of Physics, Yale University, New Haven, CT 06520, USA}
\affil{Department of Astronomy, Yale University, New Haven, CT 06520, USA}
\author{Yang Chen}
\affil{Department of Statistics, University of Michigan, Ann Arbor, MI 48109, USA}

\begin{abstract}
We present a novel population-based Bayesian inference approach to model the average and population variance of spatial distribution of a set of observables from ensemble analysis of low signal-to-noise ratio measurements. The method consists of (1) inferring the average profile using Gaussian Processes and (2) computing the covariance of the profile observables given a set of independent variables. Our model is computationally efficient and capable of inferring average profiles of a large population size from noisy measurements, without stacking and binning data nor parameterizing the shape of the mean profile. We demonstrate the performance of our method using dark matter, gas and stellar profiles extracted from hydrodynamical cosmological simulations of galaxy formation. \textsc{Population Profile Estimator (PoPE)} is publicly available in a GitHub repository. Our new method should be useful for measuring the spatial distribution and internal structure of a variety of astrophysical systems using large astronomical surveys.
\end{abstract}


\section{Introduction} 
\label{sec:introduction}

The spatial distribution and internal structure of astronomical systems contain vast amount of information about the underlying physics that governs the formation, evolution, and fate of these systems. While astronomical data collected by large-, medium-, and small-size surveys are becoming more abundant, precise and accurate modeling is becoming more challenging. The scale and complexity of these multi-wavelength surveys are exceeding the capabilities of traditional data analysis models, hence the need for novel inference models. 

One of the key challenges of modeling the empirical data is how to account for the measurement errors of varying magnitudes \citep[i.e., ``heteroscedastic’’ errors, see also][]{Kelly:2007}. The low signal-to-noise ratio (SNR) regime hinders the ability to infer the spatial structure of a population from abundant but noisy measurements, diluting the spatial signals. Typical measurements with SNR below the detection limit are often discarded or stacked to boost the signal above the detection limit \citep[{\sl e.g.},][]{McClintock:2019DES_WL, Okabe:2019, Umetsu:2017, Mezcua:2016, Greco:2015, Bulbul:2014}. Binning and stacking can introduce selection bias \citep{Dietrich:2014}, information loss \citep{Arzner:2007}, and smearing out the signal component \citep{Kipping:2010}. While stacking amplifies the SNR of the population average properties, it suppresses intrinsic scatter of the population under study. In practice, performing a statistical inference on large astronomical datasets has become a bottleneck of traditional population- and likelihood-based approaches.

In this work, we present \textsc{Population Profile Estimator (PoPE)}, a population-based, Bayesian inference model to analyze a class of problems that are concerned with the spatial distribution or internal spatial structure of a sample of astronomical systems \citep{Conselice:2014, Diemand:2011}. Our method uses the conditional statistics of spatial profile of multiple observables assuming the individual observations are measured with errors of varying magnitude. Assuming that the conditional statistics of our observables can be described with a multivariate normal distribution, the model reduces to the conditional average profile and conditional covariance between all observables. The method consists of two steps: (1) reconstructing the average profile using non-parametric regression with Gaussian Processes and (2) estimating the the property profiles covariance given a set of independent variable. Our population-based method is computationally efficient and capable of inferring average profiles of a population from noisy measurements, without stacking and binning or parameterizing the shape of the average profile.

The structure of this paper is as follows. In Section~\ref{sec:model}, we set up our model. In Section~\ref{sec:model_performance}, we assess the performance of our model in a controlled numerical experiment and show that our model can accurately reconstruct the input average profile and the covariance matrix from low SNR measurements. In Section \ref{sec:application}, we present a novel application of our model in studying the internal structure of dark matter halos. We discuss modeling of discrete observables in Section~\ref{sec:discrete_quant} and broader applications of our model in Section~\ref{sec:cumul_diff}, respectively. Finally, we conclude this work in Section~\ref{sec:summary}.

{\sl Definitions:}  Scalar variables are in italic; and vectors are in bold italic. The only matrices in this work are $\boldcorr$, $\boldSigma$, $\boldSigmaErr$. We denote logarithm base $e$ as $\ln$ and logarithm base $10$ as $\log$. Unless otherwise noted, the reported confidence regions and error bars are $68\%$ uncertainty intervals.

\section{A Population Statistics Model} \label{sec:model}

Property profiles, denoted with $\bolds$, are the primary observables, which is a function of $\boldX$, a set of independent variables. A property profile refers to spatial distribution of an observable around a population of astronomical objects, or internal spatial observables of a population of astronomical systems. For example, $s$ can be the number density profile of galaxies around a set of galaxies, or the hot gas temperature profile of dark matter halos. The first element of $\boldX$ is always the physical or angular distance from the center of astronomical objects in question (galaxies or dark matter halos in our examples). The property profiles can also be a function of other parameters (e.g., local matter density or mass of the host halo, respectively). Our aim is to model the conditional statistics of $\bolds$ given $\boldX$; i.e., $P(\bolds\,|\,\boldX)$. $P(\bolds\,|\,\boldX)$ is modelled with a multivariate normal distribution which can be specified with a mean vector $\langle \bolds\,|\, \boldX \rangle$ and the covariance matrix ${\rm Cov}(\bolds \,|\,\boldX)$. 

We propose a two-step inference model. In the first step, we infer the mean relation per data point (i.e. $\langle s_i\,|\,\boldX_i \rangle$ for data point $i$). The second step is to estimate the conditional covariance matrix given the posterior on the mean relation inferred from the first step. In Section \ref{sec:notation}, we setup the notation used throughout this work, then in Sections \ref{sec:regression} and \ref{sec:cov_inference} we explain the two steps of our model, and the implementation computational considerations are discussed in Section \ref{sec:implimentation}.

\subsection{Notation Setup} \label{sec:notation}

We shall denote the independent variable as $\boldX$ and the dependent variable as $\bolds$. In statistics literature, $\boldX$ and $\bolds$ are also referred to as the `covariate' and the `response', respectively. $\boldX$ and $\bolds$ are $N$-dimensions and $M$-dimensions vectors, respectively. We do not observe actual value of $\bolds$, instead $\bolds$ is measured with measurement noise. The measured quantities and their noise are denoted with $\hatbolds$ and $\boldSigmaErr$, respectively. 

In the following, $\langle \bolds\,|\,\boldX \rangle$ denotes the expectation value or the mean of $\bolds$ given $\boldX$. $\given{\bolds}{\boldX}$ denotes the random variable $\bolds$ conditioned on $\boldX$. $i$ is the index over data points and $\bolds_i$ and $\langle \bolds \rangle_i$ implies $\given{\bolds_i}{\boldX_i}$ and $\langle \bolds_i\,|\,\boldX_i \rangle$, respectively. The property profile vector of an astronomical system is a random variable
\begin{equation} \label{eq:model}
    \given{\bolds_i}{\boldX_i} = \langle \bolds\,|\,\boldX_i \rangle + \boldsymbol{\epsilon}(\boldX_i),
\end{equation}
where $\boldsymbol{\epsilon}_i \equiv \boldsymbol{\epsilon}(\boldX_i)$ is a random variable described with a multivariate normal distribution with mean zero. $\boldsymbol{\epsilon}$ defines the intrinsic covariance of property profiles. The covariance can also be a function of our independent variable. 

As mentioned earlier, only a noisy version of $\bolds_i$ is observed. The observed property profile is another random variable 
\begin{equation}
    \given{\hatbolds_i}{\bolds_i} = \bolds_i + \boldsymbol{\epsilon}_{i, \rm err},
\end{equation}
where $\boldsymbol{\epsilon}_{i, \rm err}$ is the measurement noise vector, which is a random variable drawn from a multivariate normal distribution with mean zero. 

Because of multivariate normal assumption in Equation (\ref{eq:model}), our model can be fully described with two quantities: (1) conditional average profile $\langle \bolds\,|\,\boldX \rangle$, and (2) the conditional profile covariance ${\rm Cov}(\bolds\,|\,\boldX) = \boldSigma$. Our goal is to estimate these two quantities in two steps by employing a data-driven model in which the form of $\langle \bolds\,|\,\boldX \rangle$ is not specified and the constraints on the amplitude, shape, and its dependency on the independent variables are inferred from the data.

\begin{table*}
\caption{Notations. } \label{tab:notation}
	\begin{center}
		\tabcolsep=0.8mm
		\begin{tabular}{ | l | l | l | }
            \hline
            Parameter & Explanation & Category \\
            \hline
            $\boldX$ & Independent variables vector. & Input variable.\\
            $\bolds$ &  Property profiles vector. & Random variable. \\
            $\hatbolds$ & Observed properties profile vector. & Random variable. \\
            \hline
            $\langle \bolds\, | \, \boldX \rangle$ & Conditional average property profiles vector, conditioned on $\boldX$. & Model Parameter. \\
            $\boldSigma$ & Conditional property profiles covariance matrix, conditioned on $\boldX$. & Model Parameter.  \\
            $\boldSigmaErr$ & Measurement error matrix. & Constant. \\
            \hline
            $l$ & Scale factor in our Gaussian Process covariance function. & Hyper-parameter. \\
            $\sigma_{\rm gp}$ & Uncertainty on the mean. & Hyper-parameter. \\
            $\sigma$ & Average population scatter for profile of a property. & Hyper-parameter.  \\
            \hline
            $i$ & Index over data points. & Index.\\ 
            $j$ & Index over the vector of property profiles. & Index. \\ 
            $k$ & Index over the vector of independent variable. & Index. \\
            \hline
            $n$ & Number of data points. & --\\ 
            $N$ & Dimension of vector $\boldX$. & -- \\ 
            $M$ & Dimension of vector $\bolds$ and $\hatbolds$. & -- \\
            \hline
    \end{tabular}
	\end{center}
\end{table*}

\subsection{Inferring the mean relation using Gaussian Process} \label{sec:regression}

The key feature of our model is employing a Gaussian process (GP) prior to reconstruct the average property profile. A large class of smooth functions can be reconstructed from a GP prior without the need of explicitly parameterizing the shape of the curves \citep{Williams:2006gp}. Its computational tractability and analytical features make it a suitable choice for our problem. A primary advantage of GPs is that it can capture non-linear and non-monotonic behaviours in the average profiles as a continuous function of the independent variables. As opposed to binning and stacking strategies, GPs enable the investigator to capture the impact of multiple variables simultaneously.

There have been surge in applications of GPs in astronomy and astrophysics data analysis, including modeling of asteroseismic data \citep{Brewer:2009}, galactic black holes light-curve \citep{Kelly:2011}, cosmic microwave background \citep{Wandelt:2003, Jewell:2004}, reconstructing the Hubble constant \citep{Melia:2018}, analyzing time domain data \citep{ForemanMackey:2017}, estimating photometric redshift \citep{Way:2009, Almosallam:2016}, and reconstructing probability distribution \citep{McClintock:2019GP}. In this work, we apply GPs to model the average profiles as a function of multiple input variables $\boldX$ in a continuous fashion and it can capture non-linear and non-monotonic trends, which is impractical with binning and stacking strategies. 

Specifically, we fit $\langle s_j \,|\,\boldX \rangle$ for each observable $j$ independently.  Without loss of generality, we drop index $j$ in the following. We employ a GP to model the mean relation as a function of the distance and other independent observes. A GP is a prior probability distribution whose domain is over the space of a continuous function \citep{Williams:2006gp}. The advantage of this model is that it does not require a parameterization of the density profile, thus is non-parametric, a binning and stacking strategy is not required, and finally due to its analytical properties the posterior has a closed form solution under minimal approximations, thereby is computationally tractable. Additionally, almost any smooth function can be reconstructed from a GP prior without the need to explicitly define the shape of the curve.   

\begin{figure*}
     \centering
     \includegraphics[width=0.98\textwidth]{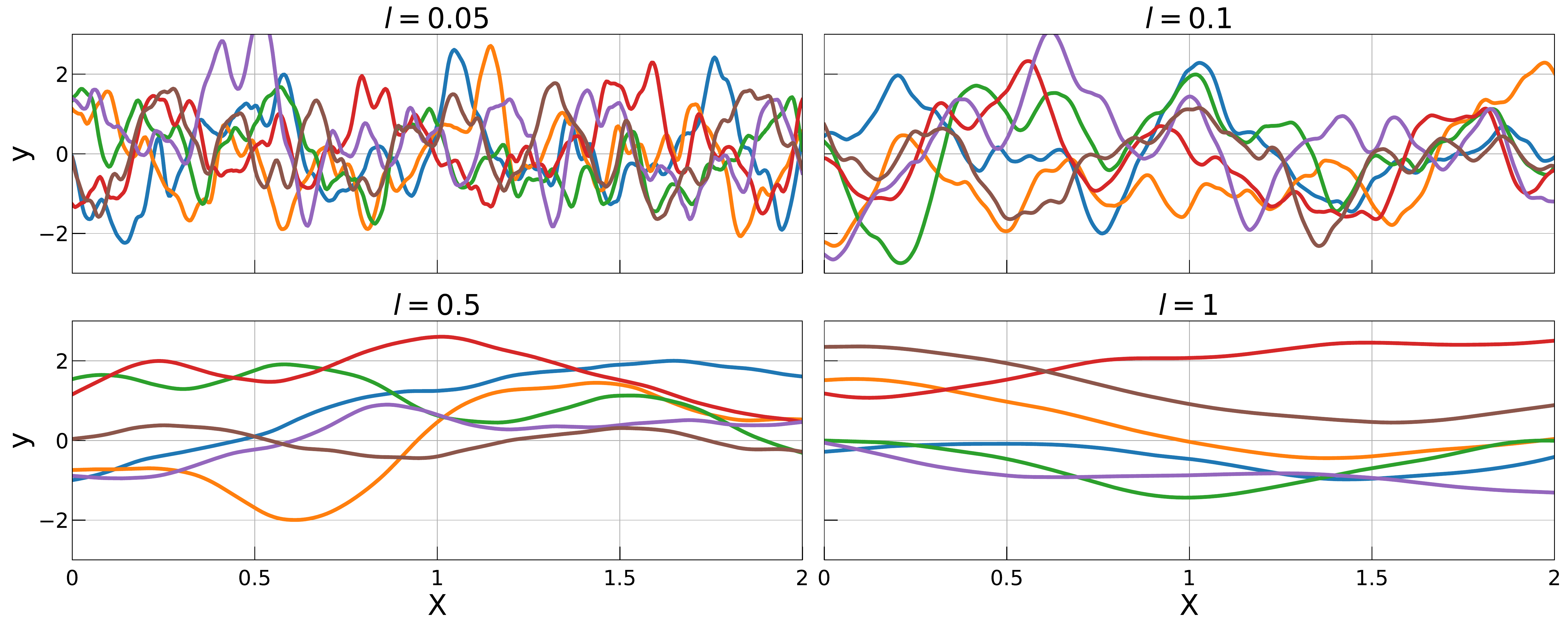} 
\caption{Example of one-dimensional curves randomly drawn from a GP prior with $m(x)=0$ and covariance function specified with a Mat\'ern kernel. The free parameters of the kernel function are set to $\sigmagp=1$ and $l \in \{0.05, 0.1, 0.5, 1\}$.}
     \label{fig:GP_examples}
\end{figure*}

A GP prior defined on the space of functions $\langle s \,|\, \boldX \rangle$
\begin{equation}
   \langle s\,|\,\boldX \rangle \sim \mathcal{G}\mathcal{P}(m(\boldX), k(\boldX, \boldX^{\prime})).
\end{equation}
The function values are modeled as a draw from a multivariate normal distribution that is parameterized by the mean function, $m(\boldX)$, and the covariance function (also known as the kernel function), $k(\boldX, \boldX^{\prime})$. Due to the marginalization and conditioning properties of the multivariate normal distribution, GPs are a convenient choice as priors over functions. Even though the goal of this work in not a prediction for a new set of observations, we note that the marginal distribution over a new point $\boldX_*$ can be evaluated easily. Here the GP prior is defined to draw the expected $s$ as a function of input variables, $\boldX$. We now need to model the property profile scatter about the conditional population average property profile. The property profile is a random variable
\begin{equation}
    \given{s}{\boldX} = \langle s\,|\,\boldX \rangle + \epsilon,
\end{equation}
where $\epsilon$ is a random variable with mean zero. We assume that $\epsilon$ is drawn from a normal distribution with variance $\sigma^2$ that is not a function of independent variables. We will revisit this assumption in Section \ref{sec:cov_inference}. Thus,
\begin{equation} 
    \given{s}{\boldX} \sim \mathcal{N}(\langle s\,|\,\boldX \rangle, \sigma^2),
\end{equation}
where $\sigma$ is a free parameter in our model.

The actual value of $s(\boldX_i)$ is not directly observable. We observe a noisy estimation of $s(\boldX_i)$ which is denoted with $\widehat{s}(\boldX_i)$. If the measurement noise has a normal distribution, $\widehat{s}(\boldX_i)$ becomes 
\begin{equation} \label{eq:gp_mean_noise}
\given{\widehat{s}}{s} \sim \mathcal{N}(s, \sigma^2_{\rm err}),
\end{equation}
where the mean value falls at the true density and the $\sigma_{\rm err}$ is the measurement uncertainty. We assume that the measurement uncertainties are known a priori. Now, we have all ingredients to build the likelihood of the observed property profile $\widehat{s}$ given $\boldX$. 

The observed profile for data point $i$ can be described with the following set of equations,
\begin{align}
    \given{\langle s \rangle_i}{\bold{x}_i} &\sim \mathcal{G}\mathcal{P}(m(\boldX_i), k(\boldX_i, \boldX_i^{\prime})), \\
    \given{s_i}{\langle s \rangle_i } &\sim \mathcal{N}(\langle s \rangle_i , \sigma^2),\\
    \given{\widehat{s}_i}{s_i} &\sim \mathcal{N}(s_i, \sigma^2_{i, \rm err}).
\end{align}
Because the marginalization over $s_i$ is analytical, the above model simplifies to
\begin{align}
    \given{\langle s \rangle_i}{\bold{x}_i} &\sim \mathcal{G}\mathcal{P}(m(\boldX_i), k(\boldX_i, \boldX_i^{\prime})), \\
    \given{\widehat{s}_i}{\langle s \rangle_i} &\sim \mathcal{N}(\langle s \rangle_i, \sigma^2_{i, \rm err}+\sigma^2). 
     \label{eq:model_mean_gp}
\end{align}

{\bf Gaussian Process Setup:} We set the mean function in the GP to zero, $m(\bold{x})=0$. Property profile can be considered as a stationary process. The best strategy to model a property profile is to re-normalize the measured profiles of a population by subtracting the average profile. We assume that a vector $\bolds$ is properly normalized, which justifies $m(\bold{x})=0$.
For the covariance, we employ the Mat\'ern kernel function \citep{Genton:2001kernels}. The class of Mat\'ern kernels is a generalization of the radial basis kernel functions (RBF) and the absolute exponential kernel functions parameterized by an additional parameter $\nu$, which specifies the smoothness of functions generated by the GP prior. We set $\nu = 5/2$, a prior over a set of two times differentiable functions. This is an important feature. In certain applications (see, Section \ref{sec:cumul_diff} for examples), we want to be able to compute the second derivative of $s$ respect to distance or log-distance. Thus, we want our function to be at least twice differentiable. The form of this kernel function is
\begin{align}
    k(\bold{x}, &\bold{x}^{\prime}) = \sigmagp^2 \times  \\ 
    & \prod_{k=1}^N  \left( 1 + \frac{\sqrt{5(x_k-x^{\prime}_k)^2}}{l_k}  + \frac{5(x_k-x^{\prime}_k)^2}{l^2_k} \right) \nonumber\\
    &  \,\,\,\,\, \times \exp\left[- \sqrt{5} \frac{(x_k-x^{\prime}_k)}{l_k}\right] . \nonumber
\end{align}
where $N$ is dimension of independent variable vector and $x_k$ is the $k^{\rm th}$ independent variable. This kernel function has two sets of free parameters, $\sigmagp$ and $l_k$. $\sigmagp$ is the overall amplitude of the GP covariance; and $l_k$ specifies the correlation scale for the $k^{\rm th}$ independent variable. We emphasize that even though the GP does not parametrize the space of potential curves, the kernel function and the mean function used to draw curves are parametric. 

For illustration, Figure~\ref{fig:GP_examples} shows several examples of one-dimensional curves randomly drawn from a GP prior for $\sigmagp=1$ and $l \in \{0.05, 0.1, 0.5, 1\}$. The scale factor $l$ sets the smoothness of function class drawn from a GP prior. A GP prior with small $l$ produces small scale fluctuations, and large $l$ produces smooth curves. If there are not enough data to constrain the small scale behaviors, small scale factor results in noisy posterior.

The kernel function $k(\boldX, \boldX^{\prime})$ is only used to define the prior distribution, and should not be confused with the covariance matrix that describes property profiles correlation in a population. In our examples, the results are insensitive to the small variation in covariance matrix hyper-parameters and even the choice of the co-variance matrix. The user needs to perform a sensitivity analysis in their application. 

At the end of inference, the only quantity which is important is the average profile at a given $\boldX$. Thus, we shall compute the posterior over $\langle s \,|\, \boldX_i \rangle$ for every single data point $i$. We perform this inference per each property profile $j$, independently. To compute the posterior distribution, we have to specify the prior for each free parameter. 

Under certain weak conditions, the posterior consistency of the GP is guaranteed \citep{choi2005posterior}, which means that as the sample size increases, the posterior distribution will concentrate around the true value of the parameter \citep{Rousseau:2016frequentist}.

{\bf Priors Setup:} Our model has in total $2 + N$ free parameters, where $N$ is the dimension of the independent variable vector. There are three categories of free parameters: (1) the scale parameters in the GP kernel function $l_k$, (2) uncertainty on the average property profile $\sigmagp$, and (3) the average population conditional scatter $\sigma$.

We use a delta function prior for $l_k$, which controls the smoothness of the GP generated curves. To keep the profile curve smooth, we do not want to set it smaller than $({\rm max}(x_k) - {\rm min}(x_k))/n$, where $n$ is the sample size. For large sample sizes, the exact value of $l_k$ become irrelevant. We set $l_k$'s prior based on the range of data, number of data points, and measurement uncertainties, that varies from one application to another. In our examples, we set $l_k = [{\rm max}(x_k) - {\rm min}(x_k)]$. We find that the final results are insensitive to variation in the $l_k$ scales. But, users need to perform a sensitivity analysis and choose a scale appropriately for their application and data.

We set a weakly informative prior on the two most important quantities in this model: $\sigmagp$ and $\sigma$, as they are physically relevant quantities. $\sigmagp$ can be interpreted as the interface uncertainty on the average profile curve given a set of data points; and $\sigma^2$ is the population average profile variance at fixed $\boldX$. We employ a half-normal distribution with large width with respect to the population variance as a prior for these two parameters. This prior is a standard choice for variance in Bayesian inference \citep[{\sl e.g.},][]{Stan:2017}.

In Equation \eqref{eq:model_mean_gp}, $\sigma$ is independent of $\boldX$ which is not strictly true and in most applications such as ours is inaccurate. Even though including a population variance in our model is necessary, the constant assumption does not have an impact on the outcome of the inferred average profile. We model the conditional dependency of the population property profile variance $\sigma$ to $\boldX$ in Section \ref{sec:cov_inference}, so we will discard $\sigma$ in our GP model after the inference.

\subsection{Inferring the covariance matrix} \label{sec:cov_inference}

Our second aim is to infer the conditional covariance matrix of a set of property profiles $\bolds$ at fixed $\boldX$. In this section, we bin, but not stack, the data and compute a conditional covariance matrix for each bin independent of other bins. In the following, we discard the bin index on $\bolds$ and $\boldX$; hence, the index on bin is implicitly assumed. The binning construction defines on which independent variables the covariance matrix is conditioned on and on which independent variables it is marginalized over. For example, if the data are binned in radius, then the estimated covariance is marginalised over halo properties and conditioned on the radial distance. An example of the binning is presented in Figure \ref{fig:SNR_low_controlled_inferred_corr} and Figure \ref{fig:TNG_mgas_mstars_corr}. In Figure \ref{fig:SNR_low_controlled_inferred_corr} data are only spatially binned, but in Figure \ref{fig:TNG_mgas_mstars_corr} bins are two-dimensional, one is spatial dimension and and the other dimension is mass of the system. 

The data consists of a vector of observed property profiles denoted with $\bolds$ (a random vector of $M$-dimension) in a given bin of $\boldX$ and their corresponding measurement error covariance. We assume that the conditional distribution of $\bolds$ given $\boldX$ is described by a multivariate Gaussian distribution
\begin{equation} \label{eq:cov_data_model}
   \given{\bolds}{\boldexps, \boldSigma} \sim  \mathcal{N}(\boldexps, \boldSigma).
\end{equation}
The mean and the covariance of this conditional distribution are denoted with a $M$-vector $\boldexps$ and a $M \times M$ matrix $\boldSigma$, respectively. We do not observe the actual values of the vector $\bolds$, but instead observe values of $\hatbolds$ which are measured with error $\boldSigmaErr$. The measured quantity is assumed to be drawn from a multivariate Gaussian distribution 
 \begin{equation}
    \given{\hatbolds}{\bolds} \sim  \mathcal{N}(\bolds, \boldSigmaErr).
 \end{equation}
We marginalize over $\bolds$ and the result is another multivariate Gaussian distribution. Each data point $i$ is generated by a multivariate Gaussian distribution
\begin{equation}
    \given{\hatbolds_i}{\boldexps_i, \boldSigma} \sim  \mathcal{N}(\boldexps_i, \boldsymbol{\Sigma}_{\rm err, i} +\boldSigma),
\end{equation}
where $i$ is the index over data. $\boldsymbol{\Sigma}_{\rm err, i}$ is the error covariance for data point $i$. The expected mean profile $\boldexps_i$ is a function of independent variables $\boldX_i$.  

Given a set of observations $\hatbolds_i$, we want to estimate the posterior distribution on the covariance matrix. The next step is to specify prior distributions for $\boldexps_i$ and $\boldSigma$. 

{\bf Setting up Priors:} To set up the priors on the mean property profiles $\boldexps_i$ per data point, we employ the posteriors estimated in Section \ref{sec:regression}. We estimate the prior on each $\boldexps_i$ with a normal distribution,
\begin{equation}
      \given{\langle s \rangle_{i, j}}{\boldX_i} \sim \mathcal{N}(\mu_{i, j}, \sigma_{\mu, i, j}) ,
\end{equation}
with mean $\mu_{i, j}$ and variance $\sigma^2_{\mu, i}$ of data point $i$ and observable $j$. $\mu_{i, j}$ and $\sigma^2_{\mu, i}$ are computed from the mean and variance of the posterior estimate of the GP model. If we ignore the uncertainty on the hyper-parameters of the GP prior, the posterior distribution on $\mu_{i, j}$ is a multivariate normal distribution. We employ the ``maximum a posteriori'' point estimation to estimate the hyper-parameters of the GP (see Section \ref{sec:implimentation} for more discussions).

We then take the diagonal components of GP posterior on $\langle s \rangle_{j}$. This implies that our prior is broader than it needs to be. In our examples and applications, the posterior variance on $\langle s \rangle_{i, j}$ is small compared to the measurement uncertainties and the intrinsic variance (see \S~\ref{sec:limitations} for more discussions). Therefore, the contribution of off-diagonal components of the prior on $\mu_{i, j}$ is negligible. Implementing the full covariance adds to the computational complexity of our model, while its impact on the posterior distribution of $\boldSigma$ is negligible. To keep the model simple and computationally efficient, we will use the diagonal components only in this work. The full covariance is a $n \times n$ matrix, where $n$ is the number of data, which becomes computationally infeasible for large data. Diagonalizing the uncertainty in the covariance matrix reduces the computational burden at a cost of inflating the uncertainty (by a few percents) on the estimated average profiles. In our applications, we show that the uncertainty on the average profiles is significantly smaller than the intrinsic scatter, and the measurement errors of the population; therefore, by a few percent over-estimation of the uncertainties of the average profile there will be inconsequential loss in the constraining power.

Constructing a proper prior for the covariance matrix is discussed in great detail in the statistics literature \citep[{\sl e.g.},][]{Barnard:2000, Lewandowski:2009generating, Alvarez:2014}. Sampling a high-dimensional covariance matrix is a difficult task, specially if the prior is not chosen carefully. We employ the \cite{Gelman:2015stan} recommended approach, which decomposes $\boldSigma$ into a correlation matrix $\boldcorr$ and a scale vector $\boldsymbol{\tau}$  \citep[see also,][]{Barnard:2000}: 
\begin{equation} \label{eq:sigmaE_decomp}
	\boldSigma = \boldvar \, \boldcorr \, \boldvar 
\end{equation}
$\boldsymbol{\tau}$ is a vector of the standard deviations of the hyper parameter $\mu$ which describe the population mean.
The prior on $\boldsymbol{\tau}$ is taken to be an inverse-Gamma distribution with shape $\alpha$ and rate $\beta$ 
\begin{equation} \label{eq:sigmaE_std_prior}
	\tau \sim {\rm inv-Gamma}(\alpha, \beta)  \propto \tau^{-2\alpha-2}\exp(-\beta \tau^{-2}).
\end{equation}
This prior is chosen so as to prevent divergences in the sampling whilst allowing large values of variance. 

\begin{figure} 
     \centering
     \includegraphics[width=0.49\textwidth]{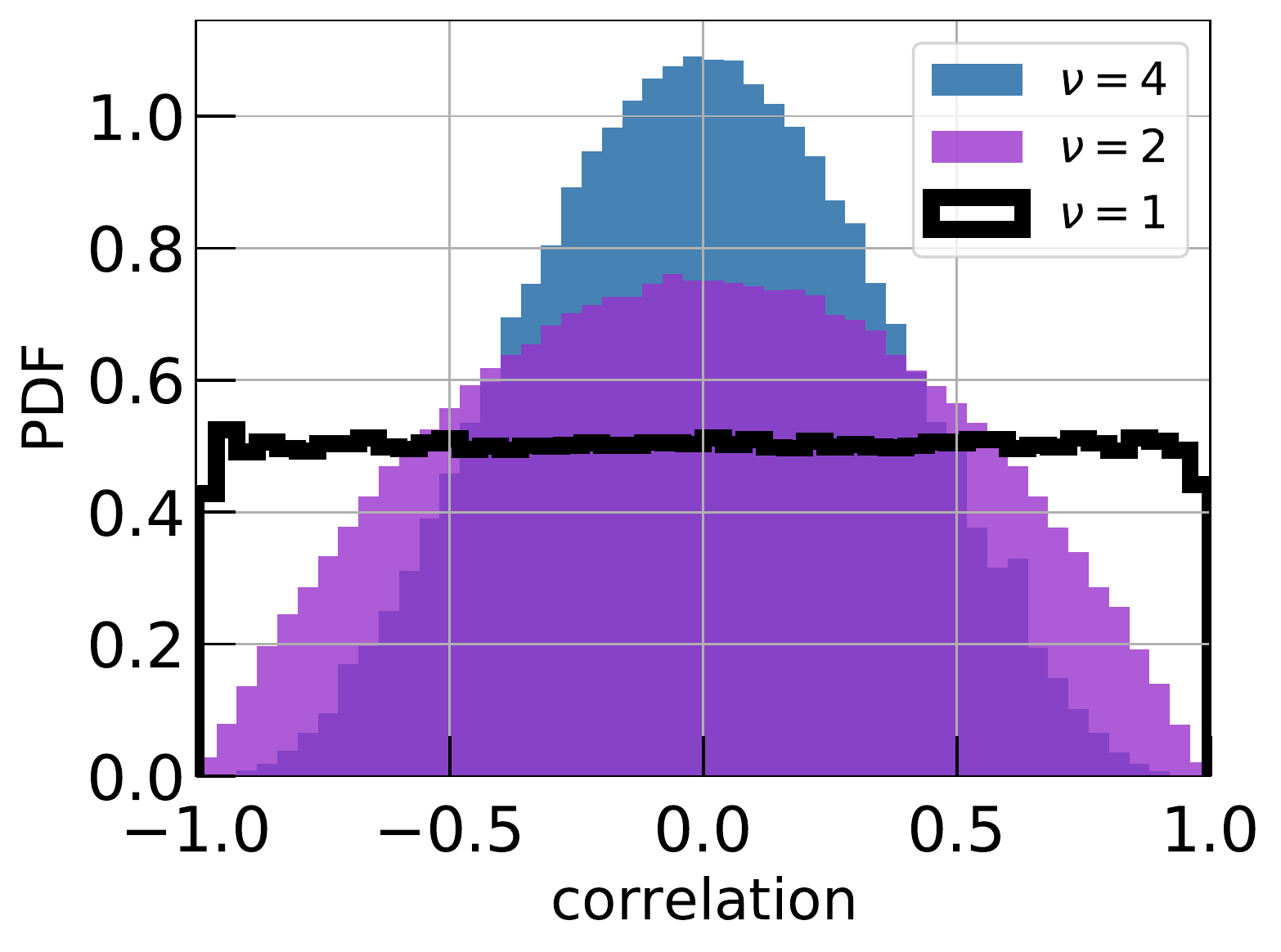}
\caption{LKJ prior distribution on the correlation coefficient of a covariance matrix of size $2\times2$ with $\nu = 1, 2,$ and $4$. }      \label{fig:LKJ_prior}
\end{figure}

An LKJ distribution prior is used on the correlation,
\begin{equation} \label{eq:sigmaE_corr_prior}
	P(\boldcorr\,|\,\nu) \propto \mathrm{det}(\boldcorr)^{\nu-1},
\end{equation}
where the shape parameter $\nu>0$. This distribution converges towards the identity matrix as $\nu$ increases (sparse covariance matrix), allowing the control of the correlation strength between the multiple parameters and consequently the variance and covariance of parameters in the population. A flat prior for marginalized elements of the correlation matrix can be imposed by setting $\nu=1$ and for $0<\nu<1$ the density has a trough at the identity matrix, which is not desired. Figure \ref{fig:LKJ_prior} compares the LKJ prior with $\nu = 1, 2,$ and $4$.

The correlation matrix $\boldsymbol{R}$ is decomposed into its Cholesky factor $\boldsymbol{L}_R$ and its transpose $\boldsymbol{L}_R^\intercal$,
\begin{align}
&	\boldcorr = \boldsymbol{L}_R \, \boldsymbol{L}_R^\intercal, \\
& 	P(\boldcorr \, | \, \nu) = \prod_{k=2}^K L_{kk}^{K-k+2\nu-2},
 \end{align}
where a LKJ prior parameterised in terms of the Cholesky decomposition has been imposed.

In our applications we do not want to force sparsity on the covariance matrix to ensure that physical models can produce large correlations between independently measured properties; hence no large values of $\nu$. Ideally the shape factor should be close to 1; i.e. marginally non-informative. We find that with $\nu = 2$ the posterior samples convergence faster than $\nu =1$ and in our toy models we recovered the input correlation. Thus, we set $\nu = 2$ that keeps the prior on the correlation matrix marginally weakly-informative. 

Rather than using the Gamma or truncated-Normal prior on the scale and the LKJ prior on the correlation, it is more common in these sorts of hierarchical analyses to set the prior on $\boldSigma$ to be the scaled inverse Wishart distribution \citep[{\sl e.g.},][]{SellentinHeavens:2016}. This choice is usually made for its analytical tractability or conjugacy on Gaussian likelihoods and simplicity within Gibbs Sampling.  However, this distribution undesirably assumes a prior relationship between the variances and correlations \citep[see][]{Alvarez:2014} which is ill-suited for our application. Our model does not have a closed-form solution, and we need to sample the posterior distribution. In a sampling method, conjugate priors are not necessary. It is worth noting that the combined scale and LKJ prior can be more efficiently sampled and gives us control over the diagonal elements of $\boldSigma$.

\subsection{Implementation and computational considerations} \label{sec:implimentation}

Main computations consist of (1) deriving the mean relation using Gaussian Process (Section~\ref{sec:regression}) and (2) computing the covariance matrix (Section~\ref{sec:cov_inference}). The pseudo-code for these two steps are provided in Algorithms \ref{alg:gp} and \ref{alg:cov}. The entire model is implemented in \texttt{PyMC3} \citep{PyMC:2016} which has an implementation of GPs, ``Fully Independent Training Conditional'' (FITC) sparse approximation, ``Maximum a Posteriori'' (MAP) estimation, and the No-U-Turn sampler (NUTS) algorithm \citep{Hoffman:2014NUTS}.

The output of the first step is the joint posterior on the average profiles for all observations, $P(\langle s \rangle_i\,|\,\hatbolds, \boldX)$. Sampling the full posterior, for large sample sizes, can become computationally infeasible and in most applications, like ours, is unnecessary. We instead estimate the posterior density employing optimization algorithms. If we set the value of $\sigmagp$ and $\sigma$, the posterior on $\langle s \rangle_i$ become analytically tractable. For a fixed $\sigmagp$ and $\sigma$, the joint posterior $P(\langle s \rangle_i\,|\, \boldX, \hatbolds)$ is a multivariate normal distribution.  We employ \texttt{find\_MAP} function in \texttt{PyMC3} for a point estimation of $\sigmagp$ and $\sigma$, that uses the Broyden-Fletcher-Goldfarb-Shanno (BFGS) optimization algorithm \citep{Fletcher:2013practical}, a fast converging, iterative method for solving unconstrained nonlinear optimization problems. We use MAP estimates to compute the GP posteriors of the average property profiles. Fixing the hyper-parameters and estimating the marginal posterior on $\langle s \rangle_i$ is sufficient for most applications.

One of the limiting factors of general GPs is their computational costs \citep{ForemanMackey:2017}. Evaluating a general GP likelihood scales as the cube of the number of data points $\mathcal{O}(n^3)$ times the dimensional of independent variables, which can become intractable even for small size data. With the current era of deep and wide surveys, we expect as many as $\sim 10^4 - 10^9$ profile measurements for a typical sample of galaxies or clusters of galaxies. For example, there are $\sim 7 \times 10^3$ optically-selected clusters in the Dark Energy Survey Year-1 data, and there are $\sim 14$ radial bins with weak lensing shear measurements \citep{McClintock:2019DES_WL}. As a result, there would be total of $\sim 10^5$ measurements. With $\mathcal{O}(n^3)$ scaling in computational costs, evaluating the posterior density is a computationally demanding task.

To address this computational bottleneck, we utilize sparse approximation methods \citet[see,][ for a review of GP sparse approximations]{Quinonero:2005unifying}. Specifically, we employ the so-called FITC approximation method \citep{Quinonero:2005unifying, Snelson:2006sparse}. This sparse approximation does not form a full covariance matrix over all $n$ data inputs. Instead it relies on a set of $m$ inducing points, where $m \ll n$.  This sparse approximation reduces the $\mathcal{O}(n^3)$ complexity of GPs down to $\mathcal{O}(nm^2)$ -- which makes the MAP estimation and average profile posterior estimation tractable. The inducing points are denoted with $\boldX_{u}$. We place these points ``uniformly'' in each dimension throughout the domain of the independent variables, $\boldX_u = x_{u,1} \wedge x_{u,N}$, where $x_{u, k} \in \{ \min(x_k), \cdots \max(x_k) \}$ and $\min(x_k)$, $\max(x_k)$ are respectively the minimum and maximum of the $k^{\rm th}$ independent variable in our sample. $m_k$, the number of inducing points on the domain of the independent variable $x_k$, is another hyper-parameter of our model that needs to be fixed. In our application, we choose $m_k$ to be between 8 and 16 points. We find that the resultant estimated posteriors on the average profiles are consistent with the input models.

We pass the estimated posterior profile mean and variance to our covariance matrix inference model. Our model does not have a closed-form solution, hence we need to sample the posterior distribution. Sampling the posterior distribution is fast enough that there is no need for an approximation. To sample the posterior distribution, we employ the NUTS algorithm \citep{Hoffman:2014NUTS}, an extension of the Hamiltonian Monte Carlo (HMC) sampling algorithm that eliminates the need to set a number of steps.

\begin{algorithm}[H]
 \caption{The conditional average property profile inference algorithm.} \label{alg:gp}
 	\begin{algorithmic}[1]
		\State \textbf{Input}: $\boldX, \widehat{s}, \sigma_{\rm err}$: a set of independent variables, observed profiles, and uncertainty of the observed profiles.
		\State \textbf{Output}: Compute posterior mean and variance of the average profile per $\boldX$.
		\vspace{0.2cm}
	    \State Set the hyper parameters $l_k$. \; 
		\State Specify priors on the hyper parameters $\sigma$ and $\sigma_{\rm gp}$. \; 
		\State Specify $\boldX_u$ array.
		\State Estimate MAP for the model parameters $\sigma$ and $\sigma_{\rm gp}$. \;
		\State Compute conditional posterior mean and variance of the average profile at $\boldX_i$ per data point $i$. \;
	\end{algorithmic} 
\end{algorithm}

\begin{algorithm}[H]
	\caption{The conditional covariance matrix inference algorithm.} \label{alg:cov}
	\begin{algorithmic}[1]
		\State \textbf{Input}: $\boldX$, $\hatbolds, \boldSigmaErr$, a set of bins, and posterior mean and variance of the average profile per $\boldX$, 
		\State \textbf{Output}: posterior samples of the covariance matrix per bin.
		\vspace{0.2cm}
		\State Construct prior on $\langle \bolds \,|\, \boldX \rangle_i $\;
		\State Initialize the hyper parameters $\nu$, $\alpha$, and $\beta$.
		\For { ${\rm bin}_i$ in $\{1, \cdots, n_{\rm bins}\}$ }
		    \State Find the subset of data in bin $i$. \;
		    \State Pass the subset to the likelihood. \;
		    \State Draw posterior samples using the NUTS algorithm. \;
		    \State Check convergence. \;
        \EndFor
	\end{algorithmic} 
\end{algorithm}

\section{Model Performance: A controlled simulation experiment} \label{sec:model_performance}

This section provides an example application of the model discussed above and assess its performance on a toy model. Our goal is to illustrate that our inference model can recover an input model. 

We start with two exponentially decaying profiles. One profile has a core and the other one is cuspy. The coefficients are arbitrary, but chose to keep the range of $\ln(\rho)$ within 2 order of magnitude. The input average profiles are
\begin{equation} \label{eq:exp_model_mean_prof_1}
 \langle \ln \rho_1(r) \rangle = -\frac{r^2}{5} - \frac{2r}{5} + 1\,,
\end{equation}
and 
\begin{equation} \label{eq:exp_model_mean_prof_2}
 \langle \ln \rho_2(r) \rangle = -\frac{r^3}{5} + \frac{2r}{5} \,.    
\end{equation}
The independent vector $\boldX$ consists of only one element $r$, thereby $N=1$; and the property profiles vector has two elements $s = \left\{ \ln(\rho_1), \ln(\rho_2) \right\}$, thereby $M=2$. We uniformly sample the independent variable $r$ 4,000 times and compute the average profiles, $\langle \ln \rho_i(x) \rangle$. We then add correlated Gaussian intrinsic noise with variance and correlation coefficient of 0.25 and 0.5, respectively. This makes a random realization of true profiles. Observed property profiles are computed by adding Gaussian noise to the true values. The width of measurement noise is random and drawn from a uniform distribution. The maximum and minimum of the uniform distribution are tuned to confine the SNR between 1 and 3. This allows us to assess the performance of our model in a low SNR regime where all observations are below a detection threshold of 5.

\begin{figure}
     \centering
     \includegraphics[width=0.49\textwidth]{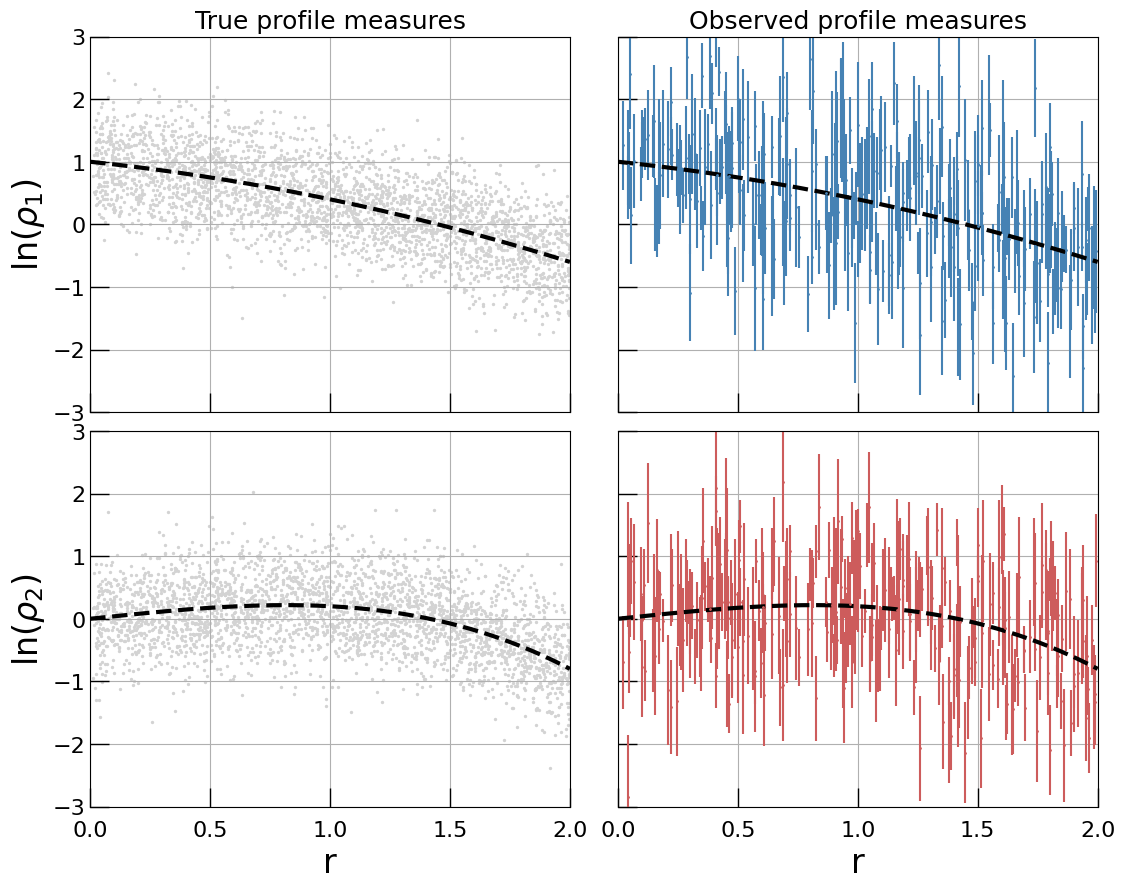} 
     \includegraphics[width=0.325\textwidth]{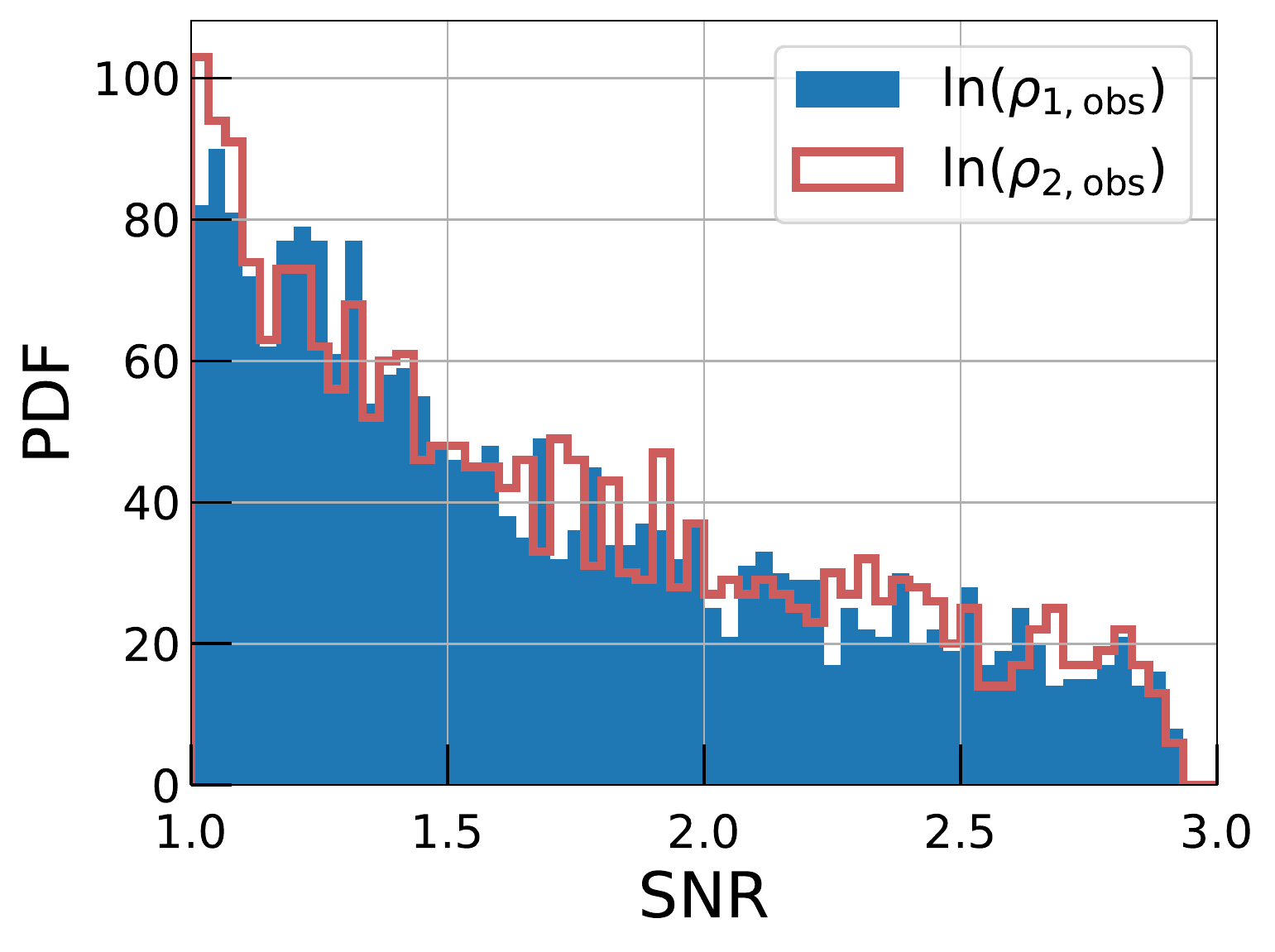} 
\caption{{\bf Top Panel:} The true average input $\ln(\rho_1)$ profile is the dashed lines, true profiles is in grey, and observed profiles are in blue with $68\%$ error bars. The true profiles are shown for the reference, the noisy measures passed into our inference model. {\bf Middle Panel:} Same as top panel but for $\ln(\rho_2)$. {\bf Bottom Panel:} The distribution of SNR for the observed $\ln(\rho_1)$ and $\ln(\rho_2)$ profiles.}
     \label{fig:SNR_low_controlled_data}
\end{figure}

Figure \ref{fig:SNR_low_controlled_data} shows a realization of the true profiles (the top two left panels) and observed profiles (the top two right panels). The input average profiles, specified in Equations (\ref{eq:exp_model_mean_prof_1}) and (\ref{eq:exp_model_mean_prof_2}),  are in black dashed lines. The true profiles are grey points in the left panels), and the observed profile measures with their $68\%$ measurement errors are shown in the right panels of Figure \ref{fig:SNR_low_controlled_data}. We pass the simulated observed profiles and their errors to our inference models and compare the inferred quantities with the input model.

\begin{figure}
     \centering
     \includegraphics[width=0.49\textwidth]{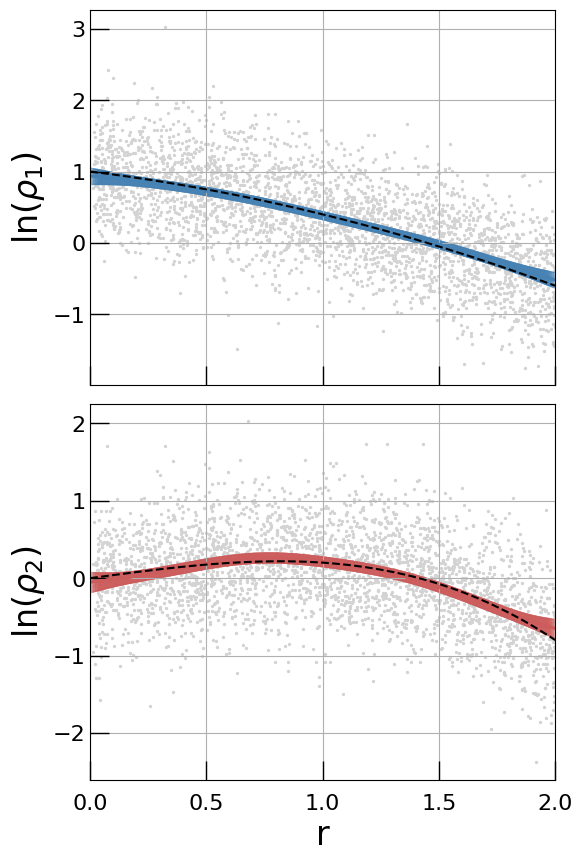} 
\caption{{\bf Top Panel:} The posterior constraints on average mean profile of $\ln(\rho_1)$ as a function of $r$. The blue region is $95\%$ posterior interval. The dashed line is the input average $\ln(\rho_1)$ profile, and grey points are true profiles. {\bf Bottom Panel:} Same as the top panel, but for $\ln(\rho_2)$. }
     \label{fig:SNR_low_controlled_inferred_mean}
\end{figure}

Figure \ref{fig:SNR_low_controlled_inferred_mean} compares the posterior average profiles (the blue and red lines) for $\ln(\rho_1)$ and $\ln(\rho_2)$ (top and bottom panels, respectively) with the input average profiles (dashed black line). The shaded regions show $68\%$ posterior intervals. The true profiles, grey points, are shown for the reference. If the posterior interval for the average profile is significantly smaller than the intrinsic scatter of the data, the intrinsic scatter and the correlations can be detected with high SNR.

\begin{figure}
     \centering
     \includegraphics[width=0.49\textwidth]{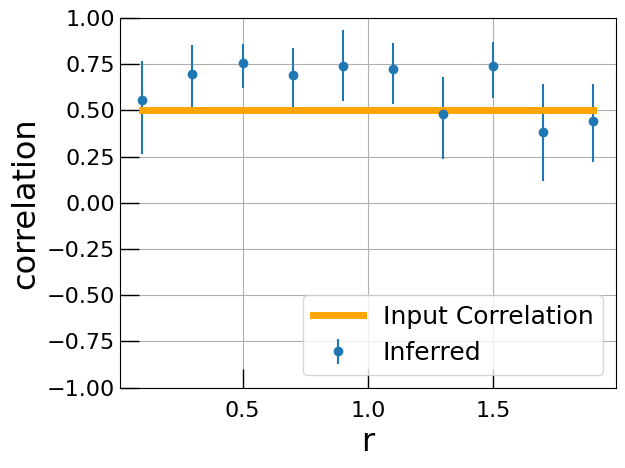} 
\caption{The correlation coefficient posterior per bin (blue points). The error bars are $68\%$ posterior intervals and the the yellow line is the input correlation coefficient. The data are binned into ten radial bins. In each bin, there are $\approx 400$ noisy data points, illustrated in Figure \ref{fig:SNR_low_controlled_data}. }
     \label{fig:SNR_low_controlled_inferred_corr}
\end{figure}

We take the output of our GP model and pass it as a prior to the covariance matrix estimator. The data are binned in 10 evenly spaced bins in $r$. There are in average 400 measured quantities in each bin. Then, we run our inference model and evaluate the posterior distribution in each bin. Figure \ref{fig:SNR_low_controlled_inferred_corr} shows the median and $68\%$ posterior intervals of the correlation coefficients (the blue points) per bin; and the yellow line is the input correlation coefficient. This figure shows that with 400 measurements of property profiles with SNR $< 3$ we can put reasonable constraints on the correlation coefficient.

\section{An application to astronomical data analysis} \label{sec:application}

One of the major challenges in the era of large multi-wavelength astronomical surveys lies in accurately modeling the spatial structures of astronomical systems for a broad mass, redshift and radial ranges \citep[e.g.,][]{Battaglia:2019,Shirasaki:2020,Salcedo:2020}.
To maximize the scientific returns of these surveys, it is important to develop a novel technique to measure the density profiles of dark matter, gas and stars and their coupling strength, which in turn contain rich information about the baryonic physics and its impact on the structure of dark matter halos \citep[e.g.,][]{Gnedin:2004,Nagai:2007,Duffy2010,Schneider2015,Schaller2015,Cui:2017,Schneider2019,Li:2020,Song2020}. Here we employ simulations to perform a forecast. We add a measurement noise to true quantities measured from simulations and show that the correlation between density profiles indeed can be estimated using low SNR measurements. 

{\bf Sample:} We employ gas and stellar density profiles derived from the TNG-100 solution of the IllustrisTNG project \footnote{\url{http://www.tng-project.org/data/}}. The simulations outputs are provided publicly by the IllustrisTNG team \citep{ Nelson:2018, Pillepich:2018, Springel:2018}. 

\begin{figure}
     \centering
     \includegraphics[width=0.49\textwidth]{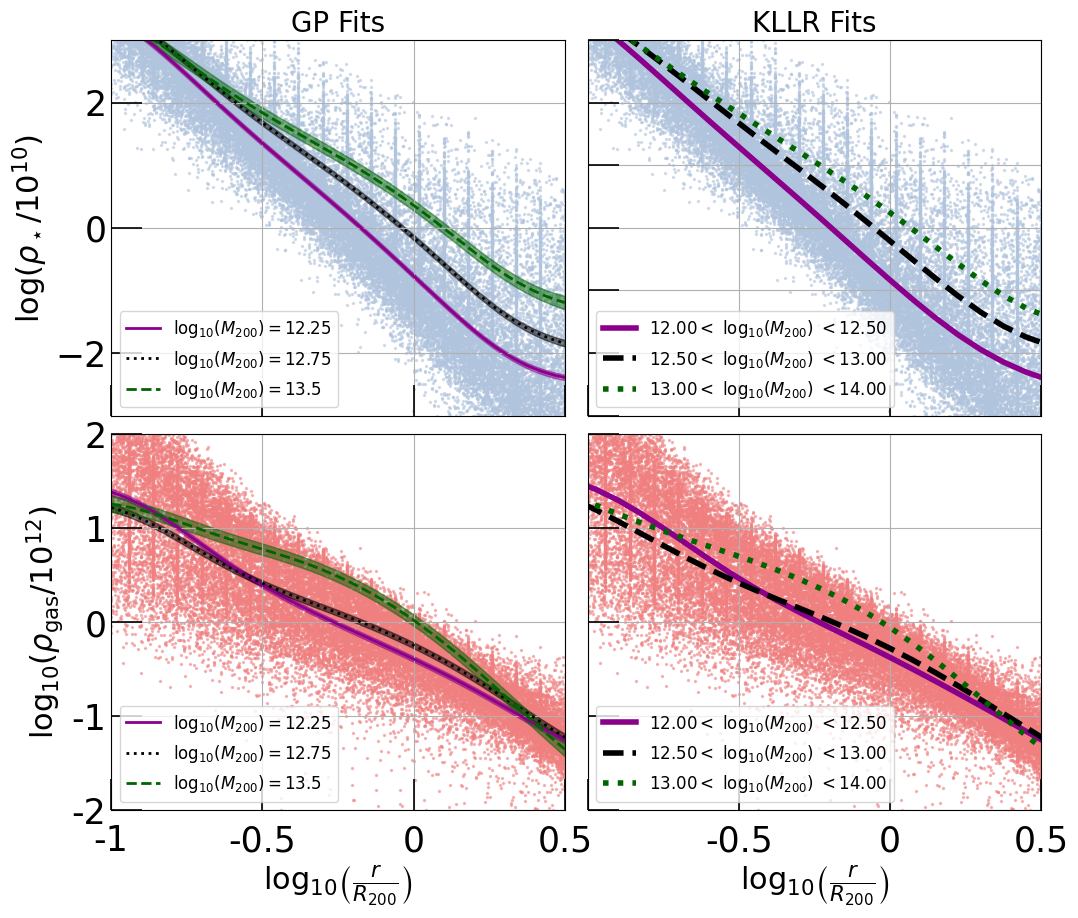} 
\caption{{\bf Top Panels:} The average inferred $\log_{10}(\rho_{\rm gas})$ profile using our GP model (left panel) and \textsc{KLLR} fit (right panel). The blue points are the true density profiles employed to estimate the \textsc{KLLR} fit, while noisy measured (not shown here) are passed to out GP model. The shaded regions in the left panel are the $68\%$ posteriors, but they are barely distinguishable from the posterior average lines. {\bf Bottom Panel:} Same at top panels but for $\log(\rho_{\star})$.}
     \label{fig:TNG_mgas_mstars_mean_profile}
\end{figure}

\begin{figure}
     \centering
     \includegraphics[width=0.33\textwidth]{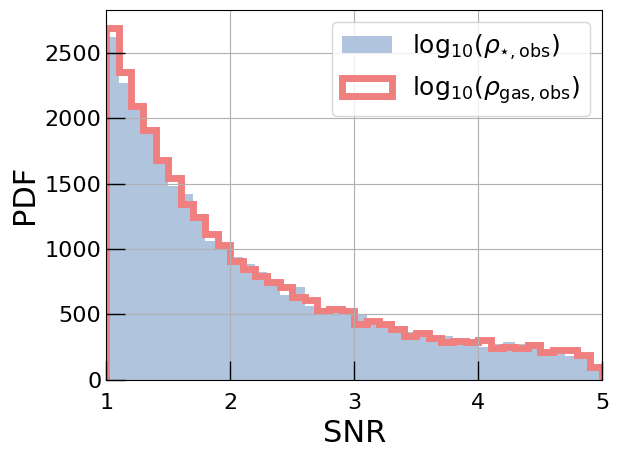} 
\caption{Signal-to-noise ratio of the mocked observed stars and gas density profile that passed to inference models. }
     \label{fig:TNG_SNR}
\end{figure}

Gas and stellar density profiles measured at $R_{200}$ normalized radius from the center of halos. Our halo sample consists of a subset of halos with mass $10^{12}\,M_{\odot}$ and above at redshift $z = 0$. Halos are identified using a ``friends-of-friends'' percolation algorithm. Our observables are gas and stellar differential density profiles. Gas and stellar density profiles are computed at different spherical shells respect to the center of a halo. The center of halos are at the minimum of the local gravitational potential. We construct a mocked observed sample by adding Gaussian noise to true density profiles (see Figures \ref{fig:TNG_mgas_mstars_mean_profile} and \ref{fig:TNG_SNR}).

 {\bf Benchmark Reference:} We fit a profile curve using a Kernel Localized Regression method motivated by \cite{Cleveland:1979robust}. Our approach is to fit a locally linear, but globally non-linear, relation to a pair of random variables $\boldX$ and $\bolds$ where $\bolds$ is a multi-dimensional random variable. We note that in our notation $\boldX$ is the independent variables and $\bolds$ is the dependent variable. To perform this regression, we employ the Kernel Localized Linear Regression (\textsc{KLLR} \href{https://github.com/afarahi/kllr}{\faGithub}) implementation motivated by \citep{Farahi:2018bahamas}. \textsc{KLLR} employs a weighted least-square fitting where the weights are assigned with a kernel (weight) function. This method allows us to model and identify non-linear behaviours of the density profile as a function of distance. The width of our Gaussian Kernel is $0.15\, [{\rm dex}]$.

\textsc{KLLR} models $\langle \boldsymbol{s}\,|\,x \rangle$ in a continuous fashion. We further bin our data into several halo mass bins to study the dependency of this conditional distribution to halo mass. We employ 1,000 bootstrap realization of the halo sample to compute the statistical uncertainty intervals. We note that the current implementation of \textsc{KLLR} cannot handle measurement uncertainties. Therefore, we can only pass true quantities to this model. We use the \textsc{KLLR} fit as the reference to compare with the output of the proposed inference model.

A limitation of the current implementation of \textsc{KLLR} is that it does not currently fit density profiles to multi-variable input data. To overcome this limitation, we bin our data into three mass bins and report the fit and correlations for halos in each bin. \textsc{KLLR} assumes that all halos in the same bin, regardless of their mass, has the same average profile. Mixing of halos in any given mass bin tends to smear out mass dependent effects and biases the resulting average profile and correlation coefficients.  The GP model, however, fits a continuous function to multi-variable input data. The bias induced by our binning in the average profile is on the order of a few percents \citep[see the mathematical derivation in][]{Evrard:2014}, and it suppresses the correlation coefficient.

{\bf Findings:} The blue and red dots in Figure \ref{fig:TNG_mgas_mstars_mean_profile} shows the true density profile of our sample; the observed quantities are not shown here, but the distribution of SNR of measured quantities is presented in Figure \ref{fig:TNG_SNR}.  Figure \ref{fig:TNG_SNR} shows the distribution of SNR for the mocked observations. We pass the mocked observations and their uncertainties to our inference model. We intentionally keep the SNR small to show that even with noisy measurements our model is capable of reconstructing the true density profile.

Figure \ref{fig:TNG_mgas_mstars_mean_profile} compares the inferred average density profiles as a function of the distance from the center of halo and their host halo mass. We compare these results by the \textsc{KLLR} fits to the true density profiles in three different mass bins. The results are presented in Figure \ref{fig:TNG_mgas_mstars_mean_profile}. The right panels are the \textsc{KLLR} fits to the true density profiles and the left panels are the GP fits to the noisy mocked density profiles. We note that the average density profiles posterior intervals are shown in Figure \ref{fig:TNG_mgas_mstars_mean_profile}, but in certain cases the 68\% posterior intervals are smaller than the width of the mean posterior lines.

\begin{figure}
     \centering
     \includegraphics[width=0.49\textwidth]{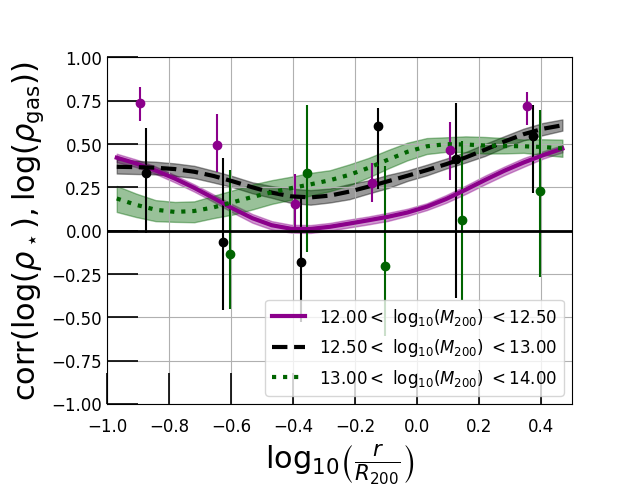} 
\caption{Correlation coefficient between the scatter of $\log(s_{\star})$ and $\log(s_{\rm gas})$ inferred from our model (from mocked noisy profiles) and the \textsc{KLLR} fits (from true simulation profiles). The error bars and shaded regions are $68\%$ confidence intervals. }
     \label{fig:TNG_mgas_mstars_corr}
\end{figure}

In Figure \ref{fig:TNG_mgas_mstars_corr}, we compare the results of inferred correlation coefficient from our model and the estimated correlation coefficient by the KLLR method. The results from two methods are in agreement. There are a few features that worth pointing out. For the most massive bin, there are only 168 halos and the SNR of mocked density profiles are pretty small, so our model cannot constrain the correlation coefficient and the posteriors are scattered around zero. For the smallest mass bin that we have a large number of halos, there is enough signal to measure the correlation coefficients and the uncertainty on the correlation coefficients from our model is pretty small. The inferred correlations from our model is consistently larger than the estimated correlation by the \textsc{KLLR} method. To compute the correlations, we need an unbiased estimation of the average profile as a function of mass and distance. Since the \textsc{KLLR} method fit a single average profile to all halos in a given mass bin, there will be additional scatter due to the mass dependency in the density profile per bin. This induced scatter suppresses the value of estimated correlation coefficients. But our GP model infers the average density profile as a function mass and radii so the correlations are not suppressed by mixing halos of different mass in a single bin. This example illustrates another advantage of our model over traditional model in which data are binned and all data in a bin treated equally. In our inference model, we still bin the data to infer the correlation, but the key in computing the covariance between two parameters is having an unbiased estimation of the average. In our inference model the averages are continuous function of mass and radii, so the estimated correlations are not suppressed in each bin.

\section{Other applications and limitations}

In this section, we review additional potential applications and limitations of our population based method.

\subsection{Derived quantities from profiles} \label{sec:cumul_diff}

It is often easier to observe the projected and cumulative profile ({\sl e.g.}, enclosed mass of halos within a sphere of radius $r$), while differential quantities are more interesting theoretically. Curves sampled from a GP can be treated like any other curve, it can be integrated, differentiated, or combined with other quantities to derive new properties. With algebraic operations, a GP posterior curves, defined on the space of observables, can be transformed to more physically relevant quantities. To compute the derived quantities, the simplest approach would be drawing curves from the posterior densities and for each curve numerically compute the quantity of interest. For instance, a set of posterior mass profiles can be differentiated and divided by the differential volumes to get posterior on differential densities. It comes with certain limitations. The GPs curves are not always differentiable, and the uncertainties grows for higher order quantities (e.g., the uncertainties on the second order derivative of a function is larger than uncertainties on the first order derivative of a function). We provide two examples. 

\begin{figure}
     \centering
     \includegraphics[width=0.49\textwidth]{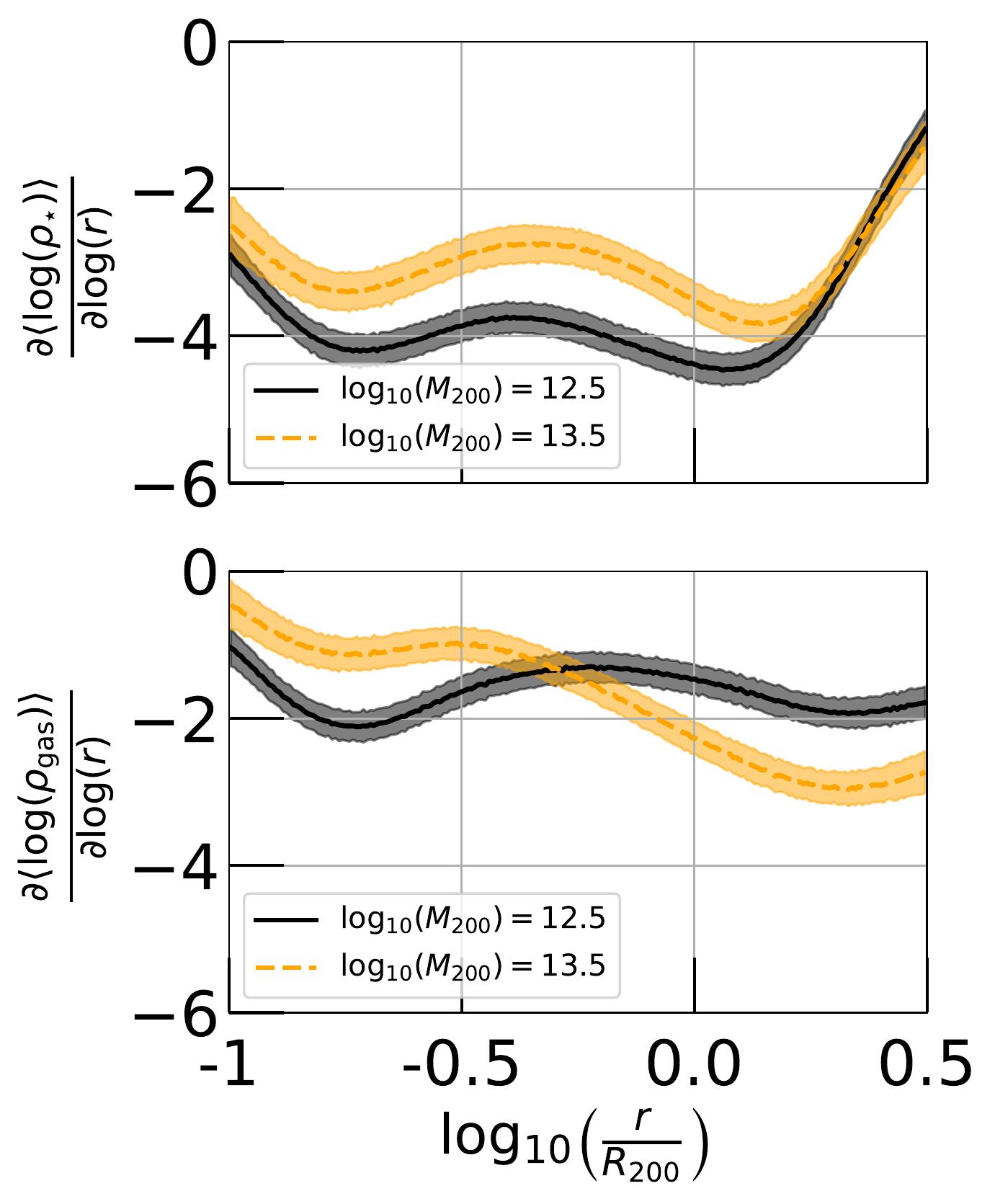} 
\caption{Inferred logarithmic slope of the gas and stellar density profile of TNG-100 halos as a function the normalized distance from the center of halos for halos of log-mass 12.5 (black solid line) and 13.5 (orange dashed line). The logarithmic slope is inferred by taking the partial derivative of $\langle \log(\rho)\rangle$ respect to $\log{r}$, where 400 random realization of $\langle \log( \rho )\rangle$ is generated from the posterior distribution of the GP model. The shaded regions are $68\%$ posterior intervals. }
     \label{fig:TNG_density_profile_log_slope}
\end{figure}

{\bf Example 1:} One of the interesting quantities of the density profiles is their logarithmic slope, $\partial\log(\rho) /\partial\log(r)$. In Section \ref{sec:application}, we inferred the average stellar and gas log-density profiles for TNG-100 halos. From the GP posters, 400 posterior average log-density profile curves are drawn. Then for each profile, the logarithmic slope is numerically computed. Figure \ref{fig:TNG_density_profile_log_slope} shows the inferred logarithmic slope of the gas and stellar density profiles. The logarithmic slopes are noisier than average profiles, but we can identify interesting features. For example the slope of stellar mass density profiles within the virial radius of halos is nearly constant and the slope for the high mass systems is shallower ($\sim -3$) than low mass end ($\sim -4$). The advantage of our GP model is that it can capture non-linear, non-monotonic trends such as the logarithmic slope of gas density profiles (see Figure \ref{fig:TNG_density_profile_log_slope}).

{\bf Example 2:} A similar approach can be applied to dwarf galaxies in order to model the logarithmic slope of their total mass density profiles. At small scales, the cold-dark matter model predicts that the inner slope of the density profile follows $\rho_{\rm dm} \propto r^{-1}$ \citep{NFW}. Empirical measurements of density profile of dwarf spheroidal galaxies suggest shallower slopes and often consistent with a constant-density core at the centre.
This disagreement between observations and simulations has become known as the core-cusp problem \citep{Bullock:2017, deBlok:2010}. With the proposed model, one can model the average density profile of a population of galaxies and study how the inner slope changes as a function of galaxy's observables. 

\subsection{Modeling Discrete Observables} \label{sec:discrete_quant}

In some applications in astronomy and astrophysics, the profile observables are discrete (e.g., the number density of halo satellite galaxies). In these applications, our noise model requires modification. Specifically, to model the measurement noise of discrete quantities, Equation~(\ref{eq:gp_mean_noise}) must be modified to 
\begin{equation} 
    \given{\widehat{s}}{\bold{x}} \sim {\rm Poisson}(\lambda = s),
\end{equation}
in order to take into account a Poisson distribution to deal with a set of discrete counting measurements, instead of a Gaussian distribution assumed earlier.

While high energy observables are promising probes of the nature of dark matter, these observables are often photon deficient \citep[{\sl e.g.},][]{DrlicaWagner:2015, Lee:2016}. A proper model of these photon deficient observations should account for noise due to small numbers in counts of photons. 

Similarly, measuring galaxies number counts in groups and clusters of galaxies is affected by discrete, small number statistics. Spatial distribution of satellite galaxies in dark matter halos is one of the key ingredients for modeling the galaxy - dark matter connection and interpreting data from large galaxy surveys \citep[{\sl e.g.},][]{Gao:2004, Nagai:2005, Sales:2007, Piscionere:2015, Agustsson:2018}. Measuring the spatial distribution of satellite galaxies per halo and inferring their density profile lies in discrete, small number statistics regime require extra care in modeling.

Another application of discrete spatial profile measurement is inferring the dark matter distribution via the distribution of stars in star clusters or nearby dwarf galaxies \citep[{\sl e.g.},][]{Merritt:1994, Nilakshi:2002, Juric:2008, Seleznev:2016, Moskowitz:2020}.

\subsection{Limitations}
\label{sec:limitations}

There are several additional sources of systematic and measurement uncertainties that must be taken into account when analyzing real datasets.

Our GP model is physics blind; i.e., we have not incorporated any physical principles into our model. While this allows to make new discoveries, it can also be a limiting factor in certain applications. For instance, inferring the dark matter density distribution can be reconstructed from the velocity distribution of stars using Jeans equations  \citep{Walker:2011}. One can incorporate physics by adding additional constraints on the relation between the GP curves for different observables. Constructing such a physical model is application dependent and is beyond the scope of this work.

Truncated data, where the selection of the data set depends on one of the dependent variables or is correlated with the dependent variables, and the data are consequently an incomplete and biased subset of a larger population \citep{Mantz:2019}. While this case can in principle be handled by modeling the selection function or imputing the missing data, we have not explicitly addressed this limitation in this work. 

Another key assumption is the fidelity of measurement uncertainties.
Our probabilistic model assumes that the measurement errors are Gaussian with zero mean and known variance. Mis-calibration of the measurement uncertainties will result in biases in both inferred intrinsic scatter and correlations. When analyzing real data, such systematic biases must be understood and controlled by developing realistic synthetic light-cone simulations for any given survey and applying the same data reduction techniques to these mock observations.

In computing the average profiles posterior, we employed only the diagonal components of GP posterior and discarded the non-diagonal contributions. In our examples, the uncertainties on the average profiles are significantly smaller than the intrinsic scatter, and the measurement errors of the population; thus such simplification is justified. However, this simplifying assumption would break down when the sample size is small, and the uncertainties on the average profiles are comparable to the intrinsic scatter. In such limit, it would be important to take full covariance matrix into account. 

\section{Summary} \label{sec:summary}

In this work, we present a population-based approach to model the average and population variance of spatial distribution of a set of observables from low signal-to-noise ratio (SNR) measurements. The first step infers the average profile given a set of independent variables and the second step models the covariance of the profile observables. Computations consist of (1) deriving the average relation using Gaussian Process (Section~\ref{sec:regression}) and (2) computing the covariance matrix (Section~\ref{sec:cov_inference}) given a set of independent variables. We illustrate the effectiveness of our model in the low SNR limit (SNR$=3$) in a controlled simulation setting as well as cosmological simulations of galaxy formation. All the codes used to produce results in this work is publicly available in a GitHub repository (\url{https://github.com/afarahi/PoPE}, \href{https://github.com/afarahi/PoPE}{\faGithubAlt}). Our new population-based method should be useful for modeling a variety of astronomical data.

\acknowledgments
This work is supported by a Michigan Institute for Data Science (MIDAS) Fellowship and Yale University. The authors would like to thank August Evrard for useful comments and feedback on the draft of this work. 

\software{PyMC3 \citep{Salvatier2016},
          Matplotlib \citep{Hunter:2007},
          KLLR.
          }

\vspace{5mm}

\bibliography{astroref}

\label{lastpage}

\end{document}